\documentclass[aps,pra,groupedaddress,twocolumn]{revtex4-1}

\usepackage{graphicx}
\usepackage{dcolumn}
\usepackage{bm}
\usepackage{amsmath}
\usepackage{amssymb}
\usepackage{latexsym}
\usepackage{epsfig}
\usepackage{amsbsy}
\usepackage{array}
\usepackage{amssymb}
\usepackage{setspace}
\usepackage{bm}
\usepackage{epstopdf}
\usepackage{subfigure}
\usepackage[colorlinks, linkcolor=blue, anchorcolor=blue, citecolor=blue, citecolor=blue,unicode=true]{hyperref}

\def\sint{\ifmmode{- \!\!\!\!\!\! \int}
    \else{\hbox{$- \!\!\!\! \int \ $}}\fi}

\begin{document}

\title{Realization and application of parity-time-symmetric oscillators in  quantum regime }

\author{Wenlin Li}
\author{Chong Li}
\author{Heshan Song}
\email{hssong@dlut.edu.cn}
\affiliation{School of Physics and Optoelectronic Engineering, Dalian University of Technology, 116024, China}
\date{\today}
\begin{abstract}
Although people have already artificially formed  parity--time ($\mathcal{PT}$) symmetry with gain and loss in a balanced manner, it is still a defect that the gain is restricted to semi--classical but not full quantum.  Here  we propose and analyze a theoretical scheme to realize full quantum oscillator $\mathcal{PT}$--symmetry.  The quantum gain is provided by a dissipation optical cavity with blue detuned laser field. After adiabatically eliminating the cavity modes, we give an effective master equation, which is a  more complete quantum description compared with non--Hermitian Hamiltonian, to reveal the quantum behaviors of such a gain oscillator. This kind of $\mathcal{PT}$--symmetry  can eliminate the dissipation effect in quantum regime. As examples, we finally apply  $\mathcal{PT}$--symmetric oscillators to enhance optomechanically induced transparency and to preserve oscillator non--classical state. 
\end{abstract}
\maketitle
\section{Introduction}  
In recent years, the notion of parity--time ($\mathcal{PT}$) symmetry has attracted considerable interest due to its potential applications in the fields of quantum optics and quantum information processing (QIP) \cite{1,2,3,4}. Since Bender and Boettcher proved that a $\mathcal{PT}$--symmetric non--Hermitian Hamiltonian ($\hat{H}^\dagger\neq \hat{H}, [\hat{H},PT]=0$) can also have real eigenvalue spectrum \cite{5,6}, realizing $\mathcal{PT}$--symmetric complex quantum systems has become a rapidly developing issue in both theoretical and experimental researches \cite{7,8,9,10,11,12,13}. Up to now, non--Hermitian--based complex quantum mechanics is still debated \cite{14}, however, attempting to test the $\mathcal{PT}$--symmetry in open quantum systems or optical systems is effective. Recent experiments have already demonstrated $\mathcal{PT}$--symmetry behaviors in a variety of physical systems \cite{15,16,17,18,19}. Among them, a simple and intuitive scheme is to link two quantum open systems with gain and loss respectively \cite{2,10,16,17,20}. As reported in Refs. \cite{2,10}, experimentalists observed that a mode splitting between two supermodes will occur with degenerate effective dissipations once the coupling intensity passes through the exceptional point (EP). Ideally, the balanced gain and loss make the eigenvalues end up on the real axis. The system supermodes perform like a closed quantum system in this case, and some environmental damage effects can be suppressed by quantum gain. So far, similar mechanism has already been applied in lots of quantum investigations, including enhancing optics nonlinearity \cite{15,21}, enhancing photon blockade \cite{16} and realizing quantum chaos \cite{17} by an extra optical gain. 

Although $\mathcal{PT}$--symmetry has made progress in optimizing QIP scheme, it is regrettable that common non--Hermitian Hamiltonian is not a complete quantum description for open gain/loss quantum systems \cite{14}. In previous works, the gains were generally introduced by classical amplification effects (e.g. parametric amplification and doped Erbium ions for waveguides and microcavities) \cite{2,10,16,17,20,21,22}. Correspondingly, a dissipation non--Hermitian Hamiltonian can be deduced by adopting Markovian quantum master equation after neglecting its the jump term or by utilizing quantum Langevin equations without input operators. However, it still remains difficult to discuss the quantum effect of such a $\mathcal{PT}$--symmetric system when our focus is not just restricted in semi--classical level \cite{16,17,20,21,22}. Realizing a gain, and then realizing a $\mathcal{PT}$--symmetric system in quantum regime, becomes a natural desire in the research field of $\mathcal{PT}$--symmetry.

In the past decade, quantized mechanical oscillators have already provided critical resources for studying basic quantum theory and QIP \cite{23,24,25,26,27}. It is well known that oscillators can constitute so--called optomechanical systems via the radiation pressure interaction between the electromagnetic and mechanical systems \cite{28}. Thanks to this radiation pressure, cavity optomechanics plays an indispensable role in the QIP scheme, and an instructive discovery is that the oscillators can be enhanced and heated (suppressed and cooled) under the blue (red) sideband by tuning the detuning of the driving fields \cite{29,30,31}. It implies that the oscillator will perform like obtaining an effective gain in this case \cite{32}, and more importantly, a quantum description for this gain can be found after eliminating the cavity field. 

In this paper, we adopt the above idea to realize $\mathcal{PT}$--symmetric oscillators in quantum regime. After adiabatically eliminating the cavity modes, we will give an effective master equation to describe the effective gain of the oscillator. For some particular quantum states (e.g. Gaussian state), this master equation is strict, which indicates some quantum effects that can not be calculated accurately by non--Hermitian Hamiltonian can be discussed perfectly in this $\mathcal{PT}$--symmetry system. In contrast to Ref. \cite{32}, here the quantum properties can be taken into account without reconsidering the eliminated system. Therefore, our $\mathcal{PT}$--symmetric oscillators can be applied in existing QIP schemes more simply and intuitively. As examples, we apply $\mathcal{PT}$--symmetric oscillators to enhance optomechanically induced transparency (OMIT) and to preserve oscillator non--classical state. We believe this novel system can provide a promising platform for QIP.

\section{Realization of oscillator gain and $\mathcal{PT}$--symmetry}
\label{Model and dynamic analysis}
\begin{figure}[]
\centering  
\includegraphics[width=3.5in]{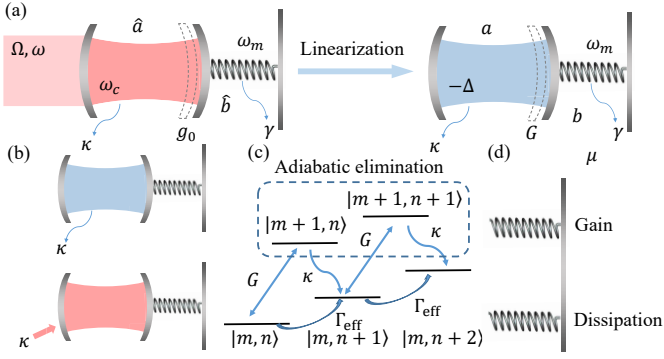}  
\caption{(a): A typical dissipative optomechanical system and its corresponding model after linearization. Such a kind of system can also be considered as a hybrid system which consists of a transmission
line resonator and a superconducting qubit. (b): Two schemes for realizing oscillator gain, corresponding to a dissipative cavity optomechanical system under blue sideband and a gain cavity optomechanical system under red sideband, respectively. (c): Level diagram of the blue sideband. Here $\vert n,m\rangle$ denotes the state of $n$ photons and $m$ phonons in the displaced frame. (d): A gain and a dissipation mutually interacting through a phonon tunneling
term of intensity $\mu$.
\label{fig:fig1}}
\end{figure}

Let us start by focusing on the realization of oscillator gain in quantum regime. As shown in Fig. \ref{fig:fig1}(a), we consider a typical optomechanical system and its the corresponding total Hamiltonian is \cite{28,29}
\begin{equation}
\begin{split}
H=-\Delta \hat{a}^{\dagger}\hat{a}+\omega \hat{b}^{\dagger}\hat{b}+g\hat{a}^{\dagger}\hat{a}(\hat{b}+\hat{b}^{\dagger})+(\Omega\hat{a}^{\dagger}+\Omega^{*}\hat{a}^{\dagger})
\label{eq:nolineHamiltonian}
\end{split}
\end{equation}
after a frame rotating.
Here $\Delta=\omega-\omega_c$ is the input cavity detuning, $\hat{a}$ ($\hat{b}$) is the annihilation operator of the optical (mechanical) mode with  the corresponding angular resonance frequency $\omega_c$ ($\omega_m$). $g$ is the single photon optomechanical coupling rate and $\Omega=\sqrt{\kappa_{ex}P/(\hbar\omega)}e^{i\phi}$ is the driving intensity with the input laser power $P$ and the initial input laser phase (cavity coupling) $\phi$ ($\kappa_{ex}$). Based on Eq. \eqref{eq:nolineHamiltonian}, the quantum Langevin equations can be expressed as:
\begin{equation}
\begin{split}
&\hat{a}=\left(i\Delta-\dfrac{\kappa}{2}\right)\hat{a}-ig\hat{a}(\hat{b}+\hat{b}^{\dagger})-i\Omega-\sqrt{\kappa}a_{in}\\
&\hat{b}=\left(-i\omega_m-\dfrac{\gamma}{2}\right)\hat{b}-ig\hat{a}^{\dagger}\hat{a}-\sqrt{\gamma}b_{in},
\label{eq:qle}
\end{split}
\end{equation}
where $\gamma$ denotes the intrinsic oscillator dissipation and $\kappa$ is the total cavity dissipation rate. If the optomechanical coupling is quite weak in quantum regime, the motion of the oscillator and optical field can be regarded as  perturbations on their respective steady states, implying that each operator in Eq. \eqref{eq:qle} can be rewritten as a sum of a $c$ number steady state value and a perturbation operator, i.e., $\hat{a}=\alpha+a$ and $\hat{b}=\beta+b$. Substituting these relations into Langevin equations and neglecting the high order perturbation terms, the system can be linearized by separating the steady states and perturbation components, and then the dynamics of perturbation operator will satisfy a linear Hamiltonian 
\begin{equation}
\begin{split}
H=-\Delta'a^{\dagger}a+\omega b^{\dagger}b+Ga^{\dagger}b^{\dagger}+G^{*}ab
\label{eq:lineHamiltonian}
\end{split}
\end{equation}
under the blue sideband condition $\Delta'\sim\omega>0$ \cite{29} (above system in Fig. \ref{fig:fig1}(b)). Here we assume $\Delta'=\Delta-2g\text{Re}(\beta)\simeq\Delta$. Unlike common linearization process in which $c$ numbers are set as the operator expectation values, here $\langle a\rangle=0$ and $\langle b\rangle=0$ are not always tenable in this case. According to Eq. \eqref{eq:lineHamiltonian}, the master equation reads:
\begin{equation}
\begin{split}
\dot{\rho}=&-i[H,\rho]\\&+\dfrac{\kappa}{2}\mathcal{L}[a]\rho+\dfrac{\gamma}{2}(n_{th}+1)\mathcal{L}[b]\rho+\dfrac{\gamma}{2}n_{th}\mathcal{L}[b^{\dagger}]\rho
\label{eq:mastereq}
\end{split}
\end{equation}
where $\mathcal{L}[o]\rho=(2o^{\dagger}\rho o-o^{\dagger}o\rho-\rho o^{\dagger}o)$ is the standard form of Lindblad superoperator. Through making premultiplications with mechanical quantity operators on both sides of Eq. \eqref{eq:mastereq}, evolutions of the system dynamics can be described by a set of partial differential equations instead of solving all elements of the density matrix $\rho$. Here we only consider first and second order mechanical quantities for convenience and the linear Hamiltonian ensures dynamic equations are closed in each order. After an iteration, the evolution equations of first and second order mechanical quantities can be simplified as (see Appendix \ref{Derivation of the oscillator effective gain} for details):
\begin{widetext}
\begin{equation}
\begin{split}
\dfrac{d}{dt}\langle b\rangle=\left[-i\left(\omega+\dfrac{4\vert G\vert^2(\Delta-\omega)}{4(\Delta-\omega)^2+{\kappa^2}}\right)-\dfrac{1}{2}\left(\gamma-\dfrac{4\vert G\vert^2\kappa}{4(\Delta-\omega)^2+{\kappa^2}}\right)\right]\langle b\rangle
\label{eq:firstorderfinnal}
\end{split}
\end{equation}
and 
\begin{equation}
\begin{split}
&\dfrac{d}{dt}\langle{b^{\dagger}b}\rangle=-\left(\gamma-\dfrac{\vert 4G^2\vert\kappa}{4(\Delta-\omega)^2+\kappa^2}\right)\langle{b^{\dagger}b}\rangle+\gamma n_{th}+\dfrac{4\vert G\vert^2\kappa}{4(\Delta-\omega)^2+\kappa^2}.
\label{eq:secondorderfinal}
\end{split}
\end{equation}
by adiabatically eliminating the optical field freedom.
\end{widetext}

The coefficients of $\langle{b}\rangle$ and $\langle{b^{\dagger}b}\rangle$ in right hands of Eqs. \eqref{eq:firstorderfinnal} and \eqref{eq:secondorderfinal} correspond to the contributions of the non--Hermitian Hamiltonian and the last two terms in Eq. \eqref{eq:secondorderfinal} are modified noise terms caused by jump operators in the master equation. Therefore, the mechanical oscillator can finally be described by the following master equation 
\begin{equation}
\begin{split}
\dot{\rho}=-i[H_{eff},\rho]+{\Gamma_{eff}}(n'_{th}+1)b\rho b^{\dagger}+{\Gamma_{eff}}n'_{th}b^{\dagger}\rho b,
\label{eq:effmaster}
\end{split}
\end{equation}
with non--Hermitian Hamiltonian 
\begin{equation}
\begin{split}
H_{eff}=(\omega_{eff}-i\dfrac{\Gamma_{eff}}{2})b^{\dagger}b,
\label{eq:e21er}
\end{split}
\end{equation}
modified thermal phonon number 
\begin{equation}
\begin{split}
n'_{th}=\dfrac{1}{\Gamma_{eff}}\left(\gamma n_{th}+\dfrac{4\vert G\vert^2\kappa}{4\Delta^2+\kappa^2}\right),
\label{eq:e2r}
\end{split}
\end{equation}
and modified initial conditions of oscillator 
\begin{equation}
\begin{split}
&\langle b\rangle'(0)\simeq\sqrt{\langle b^{\dagger}b\rangle'(0)}\\
&\langle b^{\dagger}b\rangle'(0)=\left(1+\dfrac{8\vert G\vert^2}{4(\Delta-\omega_m)^2+{\kappa^2}}\right)\langle{b^{\dagger}b}\rangle(0)\\&+\dfrac{4\vert G\vert^2\langle{a^{\dagger}a}\rangle(0)}{4(\Delta-\omega_m)^2+{\kappa^2}}-\dfrac{8G(\Delta-\omega_m)\langle{ab}\rangle(0)}{4(\Delta-\omega_m)^2+{\kappa^2}}.
\label{eq:66m}
\end{split}
\end{equation} 
Here 
\begin{equation}
\begin{split}
&\omega_{eff}=\omega+\dfrac{4\vert G^2\vert(\Delta-\omega)}{4(\Delta-\omega)^2+\kappa^2}\\
&\Gamma_{eff}=\gamma-\dfrac{4\vert G^2\vert\kappa}{4(\Delta-\omega_m)^2+\kappa^2}
\label{eq:efffpar}
\end{split}
\end{equation}
are respectively effective frequency and dissipation. We emphasize here that the only approximation used in above deduction is elimination of the optical field. Therefore, unlike non--Hermitian Hamiltonian, Eq. \eqref{eq:effmaster} contains all properties of second order expectation values of the oscillators ($\langle{b^{\dagger}b}\rangle$, $\langle{bb}\rangle$ and $\langle{b^{\dagger}b^{\dagger}}\rangle$), which means some quantum properties, for example quantum fluctuation, can also be discussed by using Eq. \eqref{eq:effmaster}. Because we only iterate the first and second order expectation value equations, Eq. \eqref{eq:effmaster} is still incomplete if it is used to solve density matrix or high order expectation values (e.g. $\langle{b^{\dagger}b}b\rangle$).

We note that the effective dissipation is reduced by the factor ${4\vert G^2\vert\kappa}/[{4(\Delta-\omega)^2+\kappa^2}]$. Physically, this is because the optical field will heat the oscillator under the blue sideband. If the linearized optomechanical coupling strength satisfies $\vert G^2\vert >\gamma[{4(\Delta-\omega)^2+\kappa^2}]/\kappa$, $\Gamma_{eff}$ will be no longer positive but a negative dissipation. The non--Hermitian Hamiltonian now describes a gain effect and the mechanical oscillations display anti-damping. 
\begin{figure}[]
\centering  
\includegraphics[width=3.5in]{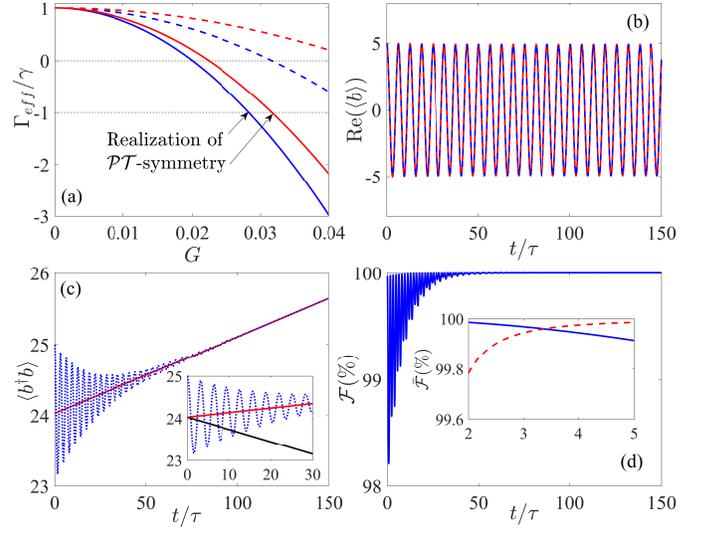}  
\caption{(a) Changes of the effective dissipation coefficient with varied linear coupling coefficient $G$. (b) and (c): Comparisons of evolutions corresponding to the first order (b) and the second order (c) mechanical quantities, respectively. Here the blue dotted lines denote mechanical quantities of original system and red solid lines denote the approximate system after the adiabatic elimination. The black line in the inset is the evolution without bath correction. (d): Fidelity between the oscillator state respectively corresponding to original and approximate systems. The inset in (d) shows the time--averaged fidelities with varied coupling coefficients $G$ (blue, solid) and the detuning $\Delta$ (red, dotted). Here the horizontal axis is $100G$ ($\Delta$) for the blue (red) line. In this simulation, we set $\omega_m=1$ as a unit and the other dimensionless parameters are: $\kappa=0.1$, $n_b=1000$. For (a) we set $\Delta=2$ ($3$) for the red (blue) lines, and $
\gamma=10^{-5}$ ($5\times 10^{-4}$) for the solid (dotted) lines. For (b), (c) and (d), we set $\gamma=10^{-5}$, $G=0.04$ and $\Delta=3$. The blue line of inset in (d) corresponds to $\Delta=3$ and the red line is under $G=0.04$.
\label{fig:fig2}}
\end{figure}

In Fig. \ref{fig:fig2}, we show the effective dissipation rates of the system under different parameters after adiabatic elimination. One can observe directly that the effective dissipation rate can appear to be negative, i.e, corresponding to a gain effect. Under some certain parameters it can be found from Fig. \ref{fig:fig2}(a) that $\Gamma_{eff}/\gamma=-1$ and the quantum dissipation is completely balanced by such a gain. In order to demonstrate the accuracy of the approximation, we plot Fig. \ref{fig:fig2}(b) and (c) to respectively compare the evolutions of the first- and second-order mechanical quantities in the approximate system and original system and it can be known that two mechanical quantities gradually exhibit  the remarkably consistent evolutions. For a Gaussian state, both the first and second order mechanical quantities are accurately described, meaning that it is a complete quantum description. In Fig. \ref{fig:fig2}(d), we plot the Gaussian fidelity to illustrate this description is complete and quantum owing to $\mathcal{F}\rightarrow 100\%$, which indicates it does not need to reconsider the eliminated system for obtaining system's quantum properties like Ref. \cite{32}, and the physical processes in our picture are more intuitive.

Let us re-examine the effective dissipation in Eq. \eqref{eq:firstorderfinnal}, 
the same conclusion can be obtained by adopting an optomechanical system with red detuning and a gain cavity (The following device in Fig. \ref{fig:fig1}(b)). However, the cavity gain is not a complete quantum description, meaning that oscillator gain in such a system is semi--classical and can not be used in quantum regime. We also emphasize that tthe necessity for correcting bath phonon number. A negative dissipation rate will change the heat flow direction between the system and the bath. Modified bath phonon number can also change the polarity of the phonons number difference between the system and the bath. Such double correction ensures the correct direction of heat flow. In the inset of Fig. \ref{fig:fig2}(c), we show system evolution corresponding to a wrong heat flow direction without correcting bath phonon number. It shows that the system should be heating are cooled, which implies only effective dissipation is not enough for the quantum description of the $\mathcal{PT}$--symmetric system.

Up to now we have discussed technique that can realize oscillator gain, generally, a $\mathcal{PT}$--symmetric system can be achieved by connecting a passive and a positive systems (see Fig. \ref{fig:fig1}(d)). The non--Hermitian Hamiltonian of such a system is:
\begin{equation}
\begin{split}
H_{eff}=&\left(\omega-i\dfrac{\gamma}{2}\right)b_1^{\dagger}b_1+\left(\omega+i\dfrac{\gamma'}{2}\right) b_2^{\dagger}b_2\\&+\mu(b_1^{\dagger}b_2+b_1b^{\dagger}_2),
\label{eq:HamiltonianPT}
\end{split}
\end{equation}
with the corresponding eigenvalues: 
\begin{equation}
\begin{split}
\lambda_{\pm}=\omega-\dfrac{i(\gamma-\gamma')}{4}\pm\sqrt{\mu^2-\left(\dfrac{\gamma+\gamma'}{4}\right)^2}.
\label{eq:eigenvalues}
\end{split}
\end{equation}
In this Hamiltonian, $\gamma$ is an ordinary oscillator dissipation and $\gamma'=-\Gamma_{eff}$ denotes a quantum gain of oscillator.

\begin{figure}[]
\centering  
\includegraphics[width=3.3in]{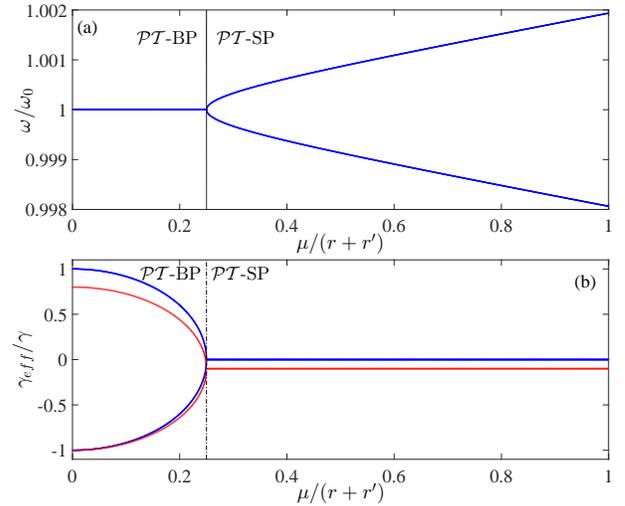}  
\caption{Real (a) and imaginary (b) parts of the eigenfrequencies of supermodes. Here blue lines in (a) and (b) respectively correspond to Re$(\lambda_{\pm})$ and Im$(\lambda_{\pm})$ as functions of coupling $\mu$. Red line in (b) denotes the system is not strict $\mathcal{PT}$--symmetry but has an effective dissipation rate.
\label{fig:fig3}}
\end{figure}

Apparently when $\mu>(\gamma+\gamma')/4$, the third term in Eq. \eqref{eq:eigenvalues} will be a pure real number. As shown in Fig. \ref{fig:fig3}, a resolved normal model splitting appears with a degenerated effective dissipation rate $\gamma-\gamma'$. The non--Hermitian Hamiltonian in this case is $\mathcal{PT}$--symmetric under the condition $\gamma=\gamma'$ and correspondingly, $\mu=(\gamma+\gamma')/4$ is exactly the exceptional point transforming from the $\mathcal{PT}$--symmetric phase ($\mathcal{PT}$SP) to $\mathcal{PT}$--symmetry breaking phase ($\mathcal{PT}$BP). Fig. \ref{fig:fig3}(b) also illustrates that the bifurcations of real and imaginary parts of the eigenvalues are still similar with the $\mathcal{PT}$SP and $\mathcal{PT}$BP for the unbalanced case in which the loss and gain parameters $\gamma$ and $\gamma'$ are unequal. This case should be regarded as physically realistic scenario while $\gamma=\gamma'$ is an idealization, i.e., a closed $\mathcal{PT}$--symmetric system is placed in hot bath with an effective dissipation rate $(\gamma-\gamma')/2$ \cite{33,34}. 

\section{applications of $\mathcal{PT}$--symmetric oscillators in quantum regime}
In this section, we present two examples of applying our $\mathcal{PT}$--symmetric oscillators in quantum regime, including enhancing optpmechanically induced transparency (\ref{Enhanced electromagnetically induced transparency}) and protecting non-classical state with the quantum gain (\ref{Protection of non classical state oscillator}). 
\subsection{Enhanced optomechanically induced transparency}
\label{Enhanced electromagnetically induced transparency}
\begin{figure}[b]
\centering  
\includegraphics[width=3.5in]{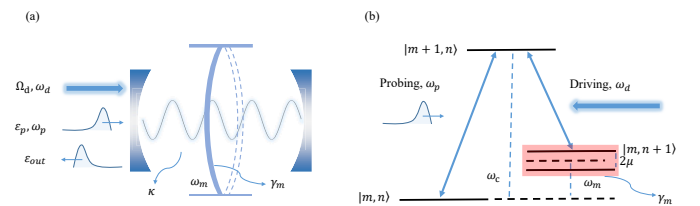}  
\caption{(a): An optical cavity coupled $\mathcal{PT}$--symmetric oscillators with nonlinear radiation pressure. (b): Level diagram of the OMIT.
\label{fig:fig4}}
\end{figure}
Electromagnetically induced transparency (EIT) is remarkable interference phenomenon in quantum optic, it has provided a promising platform for the coherent manipulation and slow light operation. In recent years, optomechanically induced transparency (OMIT) has been
widely explored both in theory and in experiment because of its well controllability \cite{22,35,36,37,38}. The underlying physics of OMIT is formally similar to that of ordinary EIT in atomic system. In Ref. \cite{35} Huang and Agarwal discussed this relation and sketched three conditions for occurring OMIT, that is, $\romannumeral1$. Driving frequency is set in the red sideband; $\romannumeral2$. Optical field loss needs to be much greater compared to that of oscillator dissipation; $\romannumeral3$. Steady state of oscillator displacement is not zero. For a normal optomechanical system, oscillator displacement is proportional to the cavity photon number, which causes that there is a conflict between the last two conditions in this case. A larger cavity dissipation satisfying $\romannumeral2$ will reduce photon number significantly, indicating that $\romannumeral3$ is violated. Fundamentally,  strong single photon couplings have to be adopted in previous works in order to display OMIT windows, but they are too difficult to realize in experiment. 

Unlike the efforts to improve the single photon coupling, we find that the requested cavity dissipation can be reduced if we use $\mathcal{PT}$--symmetric oscillators instead of dissipative oscillators in OMIT system. As shown in Fig. \ref{fig:fig4}, we consider $\mathcal{PT}$--symmetric oscillators coupled with a cavity. The total Hamiltonian of such a system can be given by the following non--Hermitian Hamiltonian in rotating framework:
\begin{widetext}
\begin{equation}
\begin{split}
H=\left(\Delta_c-i\dfrac{\kappa}{2}\right){a}^{\dagger}{a}+\sum_{j=1,2}[\omega_m b_j^{\dagger}{b}_j-g_0{a}^{\dagger}{a}({b}_j^{\dagger}+{b}_j)]+\mu({b}_1^{\dagger}{b}_2+H.c.)-i\dfrac{\gamma}{2}b_1^{\dagger}{b}_1+i\dfrac{\gamma'}{2}b_2^{\dagger}{b}_2+H_d+H_p,
\label{eq:EITHamiltonian}
\end{split}
\end{equation}
where $H_d=i\Omega_d({a}^{\dagger}-{a})$ and $H_p=i({a}^{\dagger}\varepsilon_pe^{-i\delta t}-{a}\varepsilon^{*}_pe^{i\delta t})$ respectively denote the Hamiltonian of driving and probe fields. The variables $a$, $b_1$, $b_2$, $\kappa$, $\gamma$, $\gamma'$ and $\mu$ are identically defined with previous section. $\gamma$ and $\gamma'$ are small compared to the coupling and the oscillator frequency. The oscillator Hamiltonian can be diagonalized in terms of the symmetric ($c=(b_1+b_2)/\sqrt{2}$) and antisymmetric modes ($d=(b_1-b_2)/\sqrt{2}$) as \cite{shuoming}
\begin{equation}
\begin{split}
H=\left(\Delta_c-i\dfrac{\kappa}{2}\right){a}^{\dagger}{a}+(\omega_m+\mu)c^{\dagger}c+(\omega_m-\mu)d^{\dagger}d-\sqrt{2}g_0{a}^{\dagger}{a}(c^{\dagger}+c)-i(\dfrac{\gamma-\gamma'}{2})(c^{\dagger}c+d^{\dagger}d)+H_d+H_p.
\label{eq:EITdiagHamiltonian}
\end{split}
\end{equation}
Eq. \eqref{eq:EITdiagHamiltonian} implies that the antisymmetric mode has no direct interaction with the optical mode, so that it can be further expressed as   
\begin{equation}
\begin{split}
H=\left(\Delta_c-i\dfrac{\kappa}{2}\right){a}^{\dagger}{a}+\left(\omega'_m-i\dfrac{\gamma_m}{2}\right)c^{\dagger}c-g{a}^{\dagger}{a}(c^{\dagger}+c)+H_d+H_p,
\label{eq:EITdiagHamiltoniansimple}
\end{split}
\end{equation}
here we have already set $g=\sqrt{2}g_0$, $\omega'_m=\omega_m+\mu$ and $\gamma_m=(\gamma-\gamma')/2$ for convenience. 

According to this Hamiltonian \eqref{eq:EITdiagHamiltoniansimple}, we employ the following semi--classical Langevin equations to explore the nonlinear dynamics of the system.  
\begin{equation}
\begin{split}
&\dot{q}=-\dfrac{\gamma_m}{2}q-\omega'_m p\\
&\dot{p}=-\dfrac{\gamma_m}{2}p-\omega'_m q+\sqrt{2}g\vert a\vert^2\\
&\dot{a}=\left(-i\Delta_c-\dfrac{\kappa}{2}\right)a+i\sqrt{2}gaq+\Omega_d+\varepsilon_p e^{-i\delta t}
\label{eq:omitqle}
\end{split}
\end{equation}
In Eq. \eqref{eq:omitqle}, all observable operators in quantum Langevin equations have been replaced by their expectations ($o=\langle \hat{o}\rangle$ for $o\in\{a,q,p\}$). The steady state solution of such a system can be expanded to contain many Fourier components. Under the limitation of weak strength of probe field, each operator in Eq. \eqref{eq:omitqle} will have the following form $o=o_0+o_+\varepsilon_pe^{-i\delta t}+o_-\varepsilon^{*}_pe^{i\delta t}$ by neglecting the high order terms of $\varepsilon_p$ \cite{38,39,40}. Then the optical field can be solved as (see Appendix \ref{Derivation of the absorption and dispersion spectra in OMIT} for details):
\begin{equation}
\begin{split}
a_+=\dfrac{\left\{(\omega_m^2-\delta^2-i\delta\gamma_m/2)\left[-i(\Delta+\delta)+\dfrac{\kappa}{2}\right]+\beta\right\}}{\left(\omega_m^2-\delta^2-i\delta\gamma_m/2\right)\left[i(\Delta-\delta)+\dfrac{\kappa}{2}\right]\left[-i(\Delta+\delta)+\dfrac{\kappa}{2}\right]+i2\beta\Delta},
\label{eq:solutionreslut1}
\end{split}
\end{equation}
\end{widetext}
by using the input--output relation $\varepsilon_{out}(t)+\varepsilon_pe^{-i\delta t}+\Omega_d=\gamma a$. In above expression $\beta=ig_0\omega_m x_0$ is a characteristic parameter being proportional to photon number. Similarly with previous works, we concentrate on the behaviors of $a_+$ and define $\chi=\gamma a_+$ to describe the response of the cavity optomechanical system to the probe field. According to the absorption and dispersion theory, one can determine that the real and imagery parts of $\chi$ respectively represent the behaviors of absorption and dispersion \cite{36}, and an OMIT window should satisfy Re$(\chi)\rightarrow 0$ and Im$(\chi)\rightarrow 0$ simultaneously. 
\begin{figure}[]
\centering  
\includegraphics[width=3.5in]{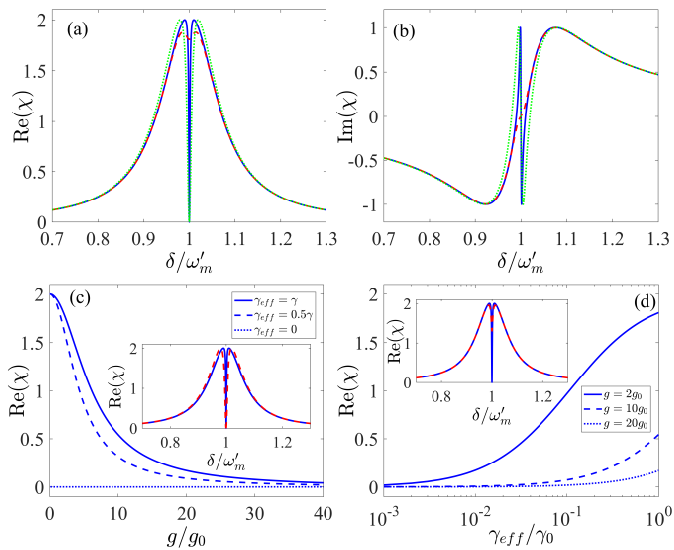}  
\caption{(a): and (b): Real and imaginary parts of $\chi$ corresponding to $\mathcal{PT}$--symmetric oscillators (blue, green) and dissipative oscillators (red), respectively. (c) and (d): Depths of transparent windows with varied single photon coupling intensity ($g$) and oscillator dissipation rate ($\gamma_m$). The insets illustrate the changes of absorption spectra under different $\gamma_{eff}$ (c) and $g$ (d). In this simulation, the parameter unit is set as $\Delta=\omega_m=1$, the other parameters are $g_0=5\times 10^{-4}$, $E=10$, $\kappa=0.15$ and $\gamma=0.02$. Oscillator gain is set as $\gamma'=0.02$ for $\mathcal{PT}$--symmetric oscillators.
\label{fig:fig5}}
\end{figure}

In Fig. \ref{fig:fig5}(a) and (b), we respectively plot the behaviors of Re$(\chi)\rightarrow 0$ and Im$(\chi)\rightarrow 0$ for both  $\mathcal{PT}$--symmetric oscillators and ordinary oscillators. It can be known that the $\mathcal{PT}$--symmetric case offers an obvious transparent window at the modified sideband $\delta=\omega_p-\omega_d=\omega_m+\mu$. This OMIT phenomenon emerges in weak coupling regime but it will not exist if the oscillator dissipation is not balanced by gain. In Fig. \ref{fig:fig5}(c) and (d) one can find that, for the similar transparent window, single-photon coupling intensity in dissipation system is required to amplify
roughly $40$ times, which is a quite harsh condition for an experiment. From this point of view, we conclude that the $\mathcal{PT}$--symmetric oscillators can indeed enhance the
OMIT phenomenon in weak coupling regime. 

\subsection{Protection of oscillator non--classical state}
\label{Protection of non classical state oscillator}
An important research significance of quantum oscillator is to prepare non--classical oscillator states in mesoscopic scale. Especially in recent years, oscillator entanglement and oscillator coherence state (e.g. Schr\"odinger Cat State) are considered as good carriers for the investigation of the boundary between classical and quantum physics \cite{24,25,26,41,42}. With the gradual deepening of the theoretical and experimental researches, people come to realize that the quantum properties of mesoscopic systems are extremely fragile in the real conditions and the coherence or entanglement is very easy to be destroyed by the complex environment \cite{41}. Therefore, a key point for realizing mesoscopic non--classical state is to suppress decoherence and to protect quantum property. From what has been discussed above, a $\mathcal{PT}$--symmetric system can be considered as a closed quantum system, and the quantum dissipation is balanced by the gain, which ensures that the quantum property can be well preserved even the oscillator with a low $Q$--factor. As an example, we discuss here how the $\mathcal{PT}$--symmetry in quantum regime can protect Gaussian entanglement between two oscillators. We re--emphasize that protecting quantum property requires a real quantum gain, however, some $\mathcal{PT}$--symmetry schemes based on classical gain can not really balance the quantum dissipation. This is the reason why $\mathcal{PT}$--symmetry is not used to protect the quantum property in previous works.

Now we go in more details. As shown in Fig. \ref{fig:fig1}(d), we consider a typical $\mathcal{PT}$--symmetric oscillator system corresponding to the non--Hermitian Hamiltonian:
\begin{equation}
\begin{split}
H=&\left(\omega-i\dfrac{\gamma}{2}\right)b_1^{\dagger}b_1+\left(\omega+i\dfrac{\gamma'}{2}\right) b_2^{\dagger}b_2\\&+\mu(b_1^{\dagger}b_2+b_1b^{\dagger}_2),
\label{eq:HamiltonianPTentna}
\end{split}
\end{equation}
and we have already proved that the expectation values of the first and second order operators satisfy the effective master equation 
\begin{equation}
\begin{split}
\dot{\rho}=&-i[H_{eff},\rho]+\gamma(n_{th}+1)b_1\rho b^{\dagger}_1+\gamma n_{th}b^{\dagger}_1\rho b_1\\&+\gamma'(n'_{th}+1)b_2\rho b^{\dagger}_2+\gamma'n'_{th}b^{\dagger}_2\rho b_2.
\label{eq:masterPTentna}
\end{split}
\end{equation}
One can utilize Eq. \eqref{eq:masterPTentna} to calculate arbitrary element in the matrix $U_{ij}=\langle \hat{u}_i\hat{u}_j\rangle$, where $\hat{u}=(\delta b^\dagger_1, \delta b_1, \delta b^\dagger_2, \delta b_2)$. And the covariance matrix $C_{ij}=\langle \hat{\xi}_i\hat{\xi}_j+\hat{\xi}_j\hat{\xi}_i\rangle/2$ for the Gaussian oscillator state can be determined by 
\begin{equation}
\begin{split}
C=\dfrac{1}{2}\left[SUS^{\top}+(SUS^{\top})^{\top}\right],
\label{eq:covariancematrix}
\end{split}
\end{equation}
where $\hat{\xi}=(\delta x_1, \delta p_1, \delta x_2, \delta p_2)$ and $S$ is the transition matrix. With the covariance matrix $C$, the entanglement between two oscillators can be measured by the \textit{logic Negativity} (see Appendix \ref{Derivation of the oscillator entanglement} for details)\cite{41,43,44} . 
\begin{figure}[]
\centering  
\includegraphics[width=3.5in]{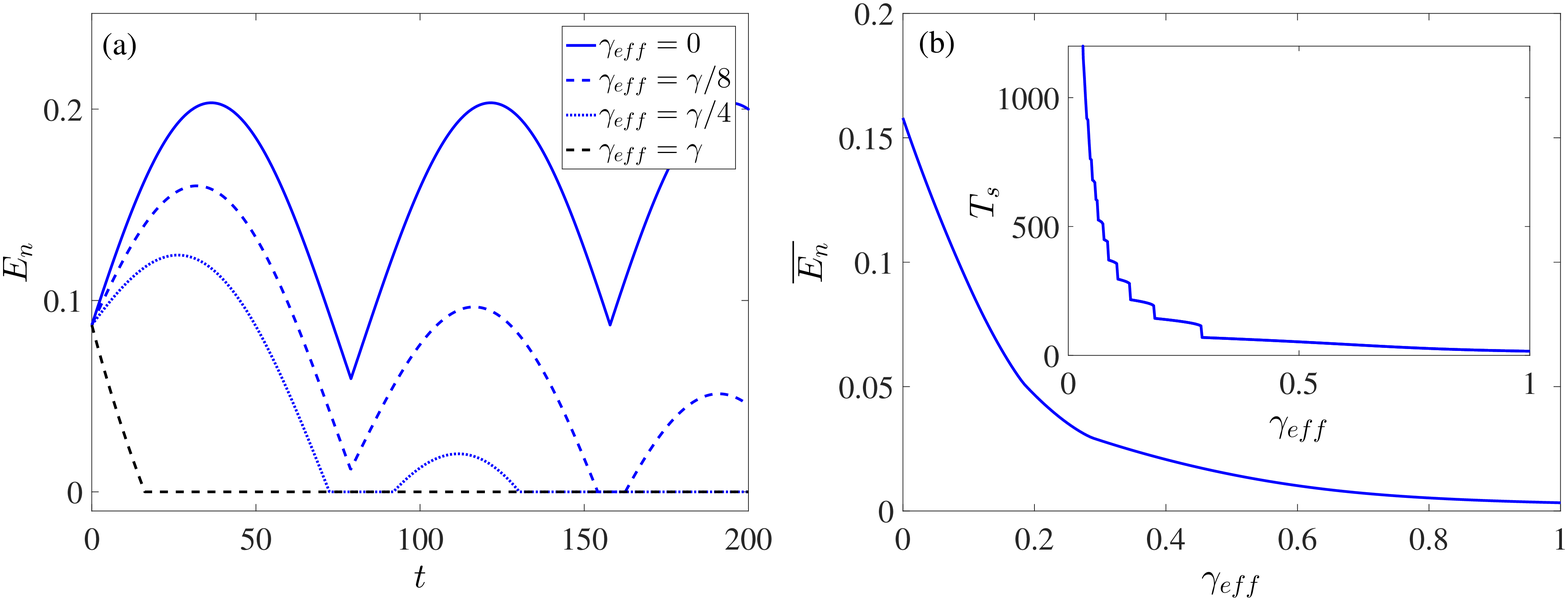}  
\caption{(a): Evolutions of \textit{logic Negativity} under different effective dissipations. (b): Time--averaged \textit{logic Negativity} with varied effective dissipation. The inset in (b) illustrates the entanglement death time $T_s$ with varied effective dissipation. In this simulation, we set $\omega=1$, $\gamma=0.004$, $\mu=0.02$ and $n_b=0$.
\label{fig:fig6}}
\end{figure}

In Fig. \ref{fig:fig6}(a), we plot the evolutions of the \textit{logic Negativity} to illustrate the protective effect of $\mathcal{PT}$--symmetry on entanglement. Here we assume that two oscillators have an initial entanglement $E_n(0)\sim 0.1$ after an entanglement preparation process. Note that the only non--local term in Hamiltonian \eqref{eq:HamiltonianPTentna} is a beam splitter (BS) type interaction, which has been proved by previous work that it can not produce steady state entanglement. Therefore for normal dissipative system ($\gamma'=-\gamma$, i.e., $\gamma_{eff}=\gamma$), entanglement will be reduced to zero immediately due to the coherence caused by the environment. With the gradually decreasing of effective dissipation, the existing entanglement time will become significantly longer. In particular at extreme case corresponding to $\gamma_{eff}=0$ ($\gamma=\gamma'$), the quantum dissipation will be completely balanced by the quantum gain and two oscillators seem to be a closed quantum system. It also can be seen from Fig. \ref{fig:fig6}(a) that the entanglement will no longer disappear in this case. 

For a quantitative description, in Fig. \ref{fig:fig6}(b), we show the time--averaged \textit{logic Negativity} in $t=200$ with the varied effective dissipation, and it can be found that there exists a monotone decreasing relation between the entanglement and the effective dissipation. When $\gamma_{eff}=0$, we observe that $\bar{E}_n=0.15$, and $\bar{E}_n>0.05$ is always satisfied even $\gamma_{eff}\sim 0.2\gamma$. In the inset of Fig. \ref{fig:fig6}(b), we show the maximum time for the nonzero entanglement, i.e, the entanglement death time $T_s$, and it can be known that  $T_s\rightarrow\infty$ when $\gamma_{eff}\rightarrow 0$, implying that for the closed-like system, the entanglement will always exist. With the increasing of effective dissipation, the death time of entanglement will reduce gradually. While $\gamma_{eff}=0.5\gamma$, we find $T_s\sim 50$ and it is still three times longer compared with that belonging to original system. 

\section{Discussions}  
\label{Discussions}
Now we give the feasibility analyses about our parameters used in above discussions. Firstly for the linearization, a well correspondence between the nonlinear Hamiltonian \eqref{eq:nolineHamiltonian} and linear Hamiltonian \eqref{eq:lineHamiltonian} has been introduced in the recent
experimental researches. Especially recent experiments have successfully realized optomechanical cooling based on the linear Hamiltonian \eqref{eq:lineHamiltonian} \cite{45,46}, and the latest theoretical work also pointed out that linearized coupling coefficient $G$ can even be controlled as a control field \cite{31}. In addition to the oscillator gain, the BS coupling between the oscillators are also widely discussed \cite{24,47,48}. Therefore, the heating effect and the $\mathcal{PT}$-symmetric scheme in our work can be easily realized by experiments. 

In the OMIT part, the dimensionless parameters are adopted according to the existing experimental parameters of OMIT and chaos in optpmechanical system, i.e., $\omega_d+\Delta_c=195$THz ($1573$nm), $\omega/2\pi=3.68$GHz, $\kappa/2\pi=500$MHz, and  $g_0/2\pi=910$kHz which correspond to $\kappa/\omega_m=0.1359\sim 0.15$ and $g_0/\omega_m=2.473\times 10^{-4}\sim 2.5\times 10^{-4}$ \cite{45,49}.   As  discussed above, our OMIT scheme does not require  a oscillator with high $Q$-factor. Therefore, some low $Q$-factor oscillator (for example the oscillators reported in Ref. \cite{51}) can also be utilized to realize OMIT. Although realizing strong opto-mechanical interaction requires to sacrifice the quality factor of the oscillator in some experiments, our scheme can alleviate this contradiction to a certain extent, which is equivalent to indirectly improving the system coupling indirectly. In the entanglement part, the relation $\gamma/\omega_m=0.004$ is also established in many recent  experiments  \cite{23,50} .

In summary, we have proposed a theoretical scheme to realize the $\mathcal{PT}$-symmetric oscillators in the quantum regime. The oscillator gain originates from an optomechanical system corresponding to a heating effect, and we can give a quantum description for such a gain by eliminating the cavity modes, which is quite different from those in the previous works in which the gain is only considered as a semi--classical amplification. Subsequently, we give an effective master equation for such a gain system containing effective dissipation, bath phonon number and initial state. This master equation allows us to obtain a complete quantum description of the $\mathcal{PT}$--symmetric oscillators, indicating that some QIP schemes can be well optimized by using our $\mathcal{PT}$--symmetric oscillator even if the semi--classical gain is powerless. As examples, we have shown how to enhance optomechanically induced transparency and how to preserve oscillator non--classical state based on the $\mathcal{PT}$-symmetric oscillators. The results respectively illustrate that Re$(\chi)\sim 0$ and Im$(\chi)\sim 0$ are easy to be satisfied even though both driving and nonlinear coupling are extremely weak, and the initial entanglement can be kept for a long time. We thus believe the scheme proposed here may provide a promising choice for the unachievable strong nonlinear coupling in quantum  optical devices and it is of potential applications for coherent manipulation, slow light operation and other utilizations in QIP.

We also note that the gain of the oscillator and the dissipative oscillator consisting of a similar closed system. This means that additional oscillator may be regarded as a special non-Markovian environment of the dissipative oscillator. In recent experiment \cite{52}, a narrow band spectral density was observed and the corresponding dynamic property is very similar to a single mode environment \cite{30}. Therefore we can predict that most QIP schemes with non-Markovian oscillator environment (e.g. Refs. \cite{30,41}) can be well extended to our $\mathcal{PT}$-symmetric systems. The possibility of the idea will be further verified in some subsequent researches.
\begin{acknowledgements}
All authors thank Jiong Cheng, Wenzhao Zhang and Yang Zhang for the useful discussion. This research was supported by the National Natural Science Foundation of China (Grant No 11175033, 11574041).
\end{acknowledgements}
\appendix

\begin{widetext}
\section{Derivation of the oscillator effective gain}
\label{Derivation of the oscillator effective gain}
Using the master equation \eqref{eq:mastereq} in the the main text, we can get the following
differential equations of first order mechanical quantities by making premultiplication with each  observable operator:
\begin{equation}
\begin{split}
&\dfrac{d}{dt}\langle a\rangle=\left(i\Delta'-\dfrac{\kappa}{2}\right)\langle a\rangle-iG\langle b\rangle^{*}\\
&\dfrac{d}{dt}\langle b\rangle=\left(-i\omega_m-\dfrac{\gamma}{2}\right)\langle b\rangle-iG\langle a\rangle^{*}.
\label{eq:firstorder}
\end{split}
\end{equation}
and correspondingly, the second-order mechanical quantities are described by:
\begin{equation}
\begin{split}
&\dfrac{d}{dt}\langle{a^{\dagger}a}\rangle=-\kappa\langle a^{\dagger}a\rangle-i[G\langle ab\rangle^{*}-G^{*}\langle ab\rangle]\\
&\dfrac{d}{dt}\langle{b^{\dagger}b}\rangle=-\gamma\langle b^{\dagger}b\rangle+\gamma n_{th}-i[G\langle ab\rangle^{*}-G^{*}\langle ab\rangle]\\
&\dfrac{d}{dt}\langle{ab}\rangle=\left[i(\Delta-\omega_m)-\dfrac{\kappa+\gamma}{2}\right]\langle ab\rangle-iG[\langle a^{\dagger}a\rangle+\langle b^{\dagger}b\rangle+1].
\label{eq:secondorder}
\end{split}
\end{equation}
Here  $\langle...\rangle$  implies taking the expectation value with respect to the density matrix of the quantum system.  Note that in the above calculations, cut-off of the density matrix is not necessary and the solutions are exact. We firstly consider the first order parts in Eq. \eqref{eq:firstorder} and these equations can be formally integrated as
\begin{equation}
\begin{split}
&\langle a\rangle(t)=\langle a\rangle(0)\exp{\left(i\Delta-\dfrac{\kappa}{2}\right)}t+\exp{\left(i\Delta t-\dfrac{\kappa}{2}t\right)}\int^{t}_0-i G\langle b\rangle^{*}(\tau)\exp{\left(-i\Delta\tau+\dfrac{\kappa}{2}\tau\right)} d\tau\\
&\langle b\rangle^{*}(t)=\langle b\rangle^{*}(0)\exp{\left(i\omega_m-\dfrac{\gamma}{2}\right)}t+\exp{\left(i\omega_mt-\dfrac{\gamma}{2}t\right)}\int^{t}_0 iG\langle a\rangle(\tau)\exp{\left(-i\omega_m\tau+\dfrac{\gamma}{2}\tau\right)} d\tau.
\label{eq:formallyintegrated}
\end{split}
\end{equation}
In the case of $\Delta-\omega_m\gg G$ or $\kappa\gg\gamma$, the optical field is a high frequency oscillation term or a highly dissipative term. Then the mode $a$ can be regarded as a perturbation on the mechanical mode. The slightly influence by $a$ can be neglected and the following approximated expression can be further obtained: 
\begin{equation}
\begin{split}
&\langle b\rangle^{*}(t)=\langle b\rangle^{*}(0)\exp{\left(i\omega_m-\dfrac{\gamma}{2}\right)}t.
\label{eq:formallysimple}
\end{split}
\end{equation}
Substituting Eq. \eqref{eq:formallysimple} into $\langle a\rangle(t)$ in Eq. \eqref{eq:formallyintegrated} and finishing the integral inside it, we can get 
\begin{equation}
\begin{split}
\langle a\rangle(t)&=\langle a\rangle(0)\exp{\left(i\Delta-\dfrac{\kappa}{2}\right)}t+\exp{\left(i\Delta t-\dfrac{\kappa}{2}t\right)}\int^{t}_0-i G\langle b\rangle^{*}(0)\exp{\left(i(\omega_m-\Delta)\tau+\dfrac{\kappa-\gamma}{2}\tau\right)} d\tau\\
&=\langle a\rangle(0)\exp{\left(i\Delta-\dfrac{\kappa}{2}\right)}t+\dfrac{-2iG}{2i(\omega_m-\Delta)+{\kappa-\gamma}}\left[\langle b\rangle^{*}(0)\exp{\left(i\omega_m-\dfrac{\gamma}{2}\right)}t-\langle b\rangle^{*}(0)\exp{\left(i\Delta-\dfrac{\kappa}{2}\right)}t\right].
\label{eq:formallyintegrateddairu}
\end{split}
\end{equation}
As we mentioned above, $\Delta-\omega_m\gg G$ or $\kappa\gg\gamma$ makes $\exp(i\Delta-\kappa/2)t$ take on a high frequency oscillation with high dissipation. Therefore, an approximate solution of the optical mode can be expressed as 
\begin{equation}
\begin{split}
\langle a\rangle(t)\simeq\dfrac{-2iG}{2i(\omega_m-\Delta)+{\kappa-\gamma}}\langle b\rangle^{*}(0)\exp{\left(i\omega_m-\dfrac{\gamma}{2}\right)}t\simeq\dfrac{-2iG}{2i(\omega_m-\Delta)+{\kappa}}\langle b\rangle^{*}(t).
\label{eq:approximatesolution}
\end{split}
\end{equation}
Substituting this approximate solution into Eq. \eqref{eq:firstorder}, we can get the following effective first order expectation value equation 
\begin{equation}
\begin{split}
\dfrac{d}{dt}\langle b\rangle&=\left[-i\left(\omega_m+\dfrac{4\vert G\vert^2(\Delta-\omega_m)}{4(\Delta-\omega_m)^2+{\kappa^2}}\right)-\dfrac{1}{2}\left(\gamma-\dfrac{4\vert G\vert^2\kappa}{4(\Delta-\omega_m)^2+{\kappa^2}}\right)\right]\langle b\rangle=\left(-i\omega_{eff}-\dfrac{\Gamma_{eff}}{2}\right)\langle b\rangle,
\label{eq:firstorderfinnalm}
\end{split}
\end{equation}
then a non-Hermitian Hamiltonian can be reversed as 
\begin{equation}
\begin{split}
H_{eff}=(\omega_{eff}-i\dfrac{\Gamma_{eff}}{2})b^{\dagger}b,
\label{eq:e21aer}
\end{split}
\end{equation}
where 
\begin{equation}
\begin{split}
&\omega_{eff}=\omega+\dfrac{4\vert G^2\vert(\Delta-\omega)}{4(\Delta-\omega)^2+\kappa^2}\\
&\Gamma_{eff}=\gamma-\dfrac{4\vert G^2\vert\kappa}{4(\Delta-\omega_m)^2+\kappa^2}
\label{eq:ejfgh}
\end{split}
\end{equation}
are exactly the same as the effective frequency and effective dissipation rate in Eq. \eqref{eq:efffpar} in the main text, respectively and Eq. \eqref{eq:firstorderfinnalm} is exactly Eq. \eqref{eq:firstorderfinnal} in the main text.

Now we consider the second-order mechanical quantities. Based on the Eq. \eqref{eq:secondorder}, we can obtain: 
\begin{equation}
\begin{split}
\langle{a^{\dagger}a}\rangle(t)=&\langle{a^{\dagger}a}\rangle(0)\exp\left(-\kappa\right)t+\exp\left(-\kappa t\right)\int^t_0-i[G\langle ab\rangle^{*}(\tau)-G^{*}\langle ab\rangle(\tau)]\exp\left(\kappa\tau\right)d\tau\\
\langle{b^{\dagger}b}\rangle(t)=&\langle b^{\dagger}b\rangle(0)\exp\left(-\gamma\right)t+\exp\left(-\gamma t\right)\int^t_0\left\{-i[G\langle ab\rangle^{*}(\tau)-G^{*}\langle ab\rangle(\tau)]+ \gamma n_{th}\right\}\exp\left(\gamma\tau\right)d\tau\\
\langle{ab}\rangle(t)=&\langle ab\rangle(0)\exp\left[i(\Delta-\omega_m)-\dfrac{\kappa+\gamma}{2}\right]t+\exp\left[i(\Delta-\omega_m)t-\left(\dfrac{\kappa+\gamma}{2}\right)t\right]\\&\times\int^t_0\{-iG[\langle a^{\dagger}a\rangle(\tau)+\langle b^{\dagger}b\rangle(\tau)+1]\}\exp\left[-i(\Delta-\omega_m)\tau+\left(\dfrac{\kappa+\gamma}{2}\right)\tau\right]d\tau,
\label{eq:seformallyintegrated}
\end{split}
\end{equation}
and the approximate solutions can also be given by:   
\begin{equation}
\begin{split}
&\langle{a^{\dagger}a}\rangle(t)=\langle{a^{\dagger}a}\rangle(0)\exp\left(-\kappa\right)t\\
&\langle{b^{\dagger}b}\rangle(t)=\langle b^{\dagger}b\rangle(0)\exp\left(-\gamma\right)t.
\end{split}
\label{eq:11}
\end{equation}
Similarly to the case of first order, and substituting Eq. \eqref{eq:11} into $\langle ab\rangle(t)$ in the above equation and finishing the integral inside it, we obtain
\begin{equation}
\begin{split}
\langle{ab}\rangle(t)=&\langle ab\rangle(0)\exp\left[i(\Delta-\omega_m)-\dfrac{\kappa+\gamma}{2}\right]t+\exp\left[i(\Delta-\omega_m)t-\left(\dfrac{\kappa+\gamma}{2}\right)t\right]\\&\times\int^t_0\{-iG[\langle{a^{\dagger}a}\rangle(0)\exp\left(-\kappa\tau\right)+\langle b^{\dagger}b\rangle(0)\exp\left(-\gamma\tau\right)+1]\}\exp\left[-i(\Delta-\omega_m)\tau+\left(\dfrac{\kappa+\gamma}{2}\right)\tau\right]d\tau\\
\simeq&-iG\left\{\dfrac{\langle{b^{\dagger}b}\rangle(t)}{{-i(\Delta-\omega_m)+\kappa/2}}+\dfrac{1}{-i(\Delta-\omega_m)+\kappa/2}\right\}+\mathcal{O}(t),
\label{eq:22}
\end{split}
\end{equation}
where
\begin{equation}
\begin{split}
\mathcal{O}(t)=&\langle ab\rangle(0)\exp\left[i(\Delta-\omega_m)-\dfrac{\kappa}{2}\right]t+\dfrac{-iG}{-i(\Delta-\omega_m)+\kappa/2}\left(-\langle{b^{\dagger}b}\rangle(0)\exp\left[i(\Delta-\omega_m)-\dfrac{\kappa}{2}\right]t\right)\\&+\dfrac{-iG}{-i(\Delta-\omega_m)-\kappa/2}\left(\langle{a^{\dagger}a}\rangle(0)\exp(-\kappa t)-\langle{a^{\dagger}a}\rangle(0)\exp\left[i(\Delta-\omega_m)-\dfrac{\kappa}{2}\right]t\right).
\label{eq:2wer2}
\end{split}
\end{equation}
We firstly neglect $\mathcal{O}(t)$ term in Eq. \eqref{eq:22} because it is a time oscillation term with dissipation, then the modified second-order equation is 
\begin{equation}
\begin{split}
\dfrac{d}{dt}\langle{b^{\dagger}b}\rangle=&-\gamma\langle b^{\dagger}b\rangle+\gamma n_{th}-i[G\langle ab\rangle^{*}-G^{*}\langle ab\rangle]\\
=&-\gamma\langle b^{\dagger}b\rangle+\dfrac{4\vert G\vert^2\kappa}{4(\Delta-\omega_m)^2+{\kappa^2}}\langle b^{\dagger}b\rangle+\gamma n_{th}+\dfrac{4\vert G\vert^2\kappa}{4(\Delta-\omega_m)^2+{\kappa^2}}\\
=&-\Gamma_{eff}\langle b^{\dagger}b\rangle+\Gamma_{eff} n'_{th}.
\label{eq:33}
\end{split}
\end{equation} 
In this expression, $\Gamma_{eff}=\gamma-(4\vert G\vert^2\kappa)/(4(\Delta-\omega_m)^2+{\kappa^2})$ is the effective dissipation and we find that this conclusion is self--consistent with the first-order effective dissipation. $n'_{th}=(\gamma n_{th}+4\vert G\vert^2\kappa/(4\Delta^2+\kappa^2))/\Gamma_{rff}$ is the effective phonon number of the bath to ensure correct direction of heat flow under the effective dissipation. Eq. 
\eqref{eq:33} is exactly Eq. \eqref{eq:secondorderfinal} in the main text. By using it, we have:
\begin{equation}
\begin{split}
\langle{b^{\dagger}b}\rangle(t)=&\langle b^{\dagger}b\rangle(0)\exp\left(-\Gamma_{eff}\right)t+\exp\left(-\Gamma_{eff} t\right)\int^t_0\left\{-i[G\mathcal{O}^{*}(\tau)-G^{*}\mathcal{O}]+ \Gamma_{eff} n'_{th}\right\}\exp\left(\Gamma_{eff}\tau\right)d\tau.
\label{eq:44}
\end{split}
\end{equation} 
After completing integration in the above expression, we achieve the following relationships:
\begin{equation}
\begin{split}
\exp\left(-\Gamma_{eff} t\right)\int^t_0\left\{\dfrac{-4\vert G\vert^2\kappa\langle{a^{\dagger}a}\rangle(0)\exp(-\kappa \tau)}{4(\Delta-\omega_m)^2+{\kappa^2}}\right\}\exp\left(\Gamma_{eff}\tau\right)d\tau=\mathcal{Z}_1-\left(\dfrac{4\vert G\vert^2\langle{a^{\dagger}a}\rangle(0)}{4(\Delta-\omega_m)^2+{\kappa^2}}\right)\exp\left(-\Gamma_{eff} t\right),
\label{eq:x1}
\end{split}
\end{equation} 
\begin{equation}
\begin{split}
&\langle{a^{\dagger}a}\rangle(0)\exp\left(-\Gamma_{eff} t\right)\int^t_0\left\{-\dfrac{\vert G\vert^2\exp\left[-i(\Delta-\omega_m)-\dfrac{\kappa}{2}\right]t}{i(\Delta-\omega_m)-\dfrac{\kappa}{2}}-\dfrac{\vert G\vert^2\exp\left[i(\Delta-\omega_m)-\dfrac{\kappa}{2}\right]t}{-i(\Delta-\omega_m)-\dfrac{\kappa}{2}}\right\}\exp\left(\Gamma_{eff}\tau\right)d\tau\\=&\mathcal{Z}_2+2\left(\dfrac{4\vert G\vert^2\langle{a^{\dagger}a}\rangle(0)}{4(\Delta-\omega_m)^2+{\kappa^2}}\right)\exp\left(-\Gamma_{eff} t\right),
\label{eq:x2}
\end{split}
\end{equation} 
\begin{equation}
\begin{split}
&\exp\left(-\Gamma_{eff} t\right)\\&\times\int^t_0\left\{-iG\langle{ab}\rangle(0)^{*}\exp\left[-i(\Delta-\omega_m)-\dfrac{\kappa}{2}\right]\tau+iG^{*}\langle{ab}\rangle(0)\exp\left[i(\Delta-\omega_m)-\dfrac{\kappa}{2}\right]\tau\right\}\exp\left(\Gamma_{eff}\tau\right)d\tau\\=&\mathcal{Z}_3-\left(\dfrac{2\text{Re}(G\langle{ab}\rangle(0))+\text{Im}(G\kappa))(\Delta-\omega_m)}{(\Delta-\omega_m)^2+\kappa^2/4}\right)\exp\left(-\Gamma_{eff} t\right)=\mathcal{Z}_3-2\left(\dfrac{4G(\Delta-\omega_m)\langle{ab}\rangle(0)}{4(\Delta-\omega_m)^2+{\kappa^2}}\right)\exp\left(-\Gamma_{eff} t\right),
\label{eq:x3}
\end{split}
\end{equation} 
and 
\begin{equation}
\begin{split}
&\langle{b^{\dagger}b}\rangle(0)\exp\left(-\Gamma_{eff} t\right)\int^t_0\left\{-\dfrac{\vert G\vert^2\exp\left[-i(\Delta-\omega_m)-\dfrac{\kappa}{2}\right]\tau}{i(\Delta-\omega_m)+\dfrac{\kappa}{2}}-\dfrac{\vert G\vert^2\exp\left[i(\Delta-\omega_m)-\dfrac{\kappa}{2}\right]\tau}{-i(\Delta-\omega_m)+\dfrac{\kappa}{2}}\right\}\exp\left(\Gamma_{eff}\tau\right)d\tau\\=&\mathcal{Z}_4+2\vert G\vert^2\langle{b^{\dagger}b}\rangle(0)\left(\dfrac{(\Delta-\omega_m)^2-\kappa^2/4}{\left[(\Delta-\omega_m)^2+\kappa^2/4\right]^2}\right)\exp\left(-\Gamma_{eff} t\right)\simeq \mathcal{Z}_4+2\left(\dfrac{4\vert G\vert^2\langle{b^{\dagger}b}\rangle(0)}{4(\Delta-\omega_m)^2+\kappa^2}\right)\exp\left(-\Gamma_{eff} t\right).
\label{eq:x4}
\end{split}
\end{equation} 
Here the final equal relationship in Eq. \eqref{eq:x3} requires $G$ and $\langle{ab}\rangle(0)$ are both real, and the last approximation in Eq. \eqref{eq:x4} needs $\kappa^2/4\ll (\Delta-\omega_m)^2$. In Eqs. \eqref{eq:x1}$\sim$\eqref{eq:x4}, $\mathcal{Z}_{1,2,3,4}$ are the terms corresponding to integral upper limit, the oscillation terms $\exp[\pm i(\Delta-\omega_m)t]$ and dissipation terms $\exp(-\kappa t/2)$ in them allow us to neglect them. Then substituting  Eqs. \eqref{eq:x1}$\sim$\eqref{eq:x4} into Eq. \eqref{eq:44}, we obtain:
\begin{equation}
\begin{split}
\langle{b^{\dagger}b}\rangle(t)=& \langle b^{\dagger}b\rangle'(0)\exp\left(-\Gamma_{eff}\right)t+\exp\left(-\Gamma_{eff} t\right)\int^t_0\left\{\Gamma_{eff} n'_{th}\right\}\exp\left(\Gamma_{eff}\tau\right)d\tau,
\label{eq:55}
\end{split}
\end{equation} 
with 
\begin{equation}
\begin{split}
\langle b^{\dagger}b\rangle'(0)=\left(1+\dfrac{8\vert G\vert^2}{4(\Delta-\omega_m)^2+{\kappa^2}}\right)\langle{b^{\dagger}b}\rangle(0)+\dfrac{4\vert G\vert^2\langle{a^{\dagger}a}\rangle(0)}{4(\Delta-\omega_m)^2+{\kappa^2}}-\dfrac{8G(\Delta-\omega_m)\langle{ab}\rangle(0)}{4(\Delta-\omega_m)^2+{\kappa^2}}
\label{eq:66}
\end{split}
\end{equation} 
can be considered as a modified initial condition of the oscillator phonon number. And correspondingly, we assume $\langle b\rangle'(0)=\sqrt{\langle b^{\dagger}b\rangle'(0)}$ for convenience. Eq. \eqref{eq:66} is exactly the same with the modified initial condition in Eq. \eqref{eq:66m} in the main text.

\section{Derivation of the absorption and dispersion spectra in OMIT}
\label{Derivation of the absorption and dispersion spectra in OMIT}
Substituting relation $o=o_0+o_+\varepsilon_pe^{-i\delta t}+o_-\varepsilon^{*}_pe^{i\delta t}$ into semi--classical Langevin equations (Eq. \eqref{eq:omitqle} in the main text) and using the steady state condition, we can give the dynamic equations of systemic mean values as: 
\begin{equation}
\begin{split}
&0=-\dfrac{\gamma_m}{2}q_0-\omega'_m p_0\\
&0=-\dfrac{\gamma_m}{2}p_0-\omega'_m q_0+\sqrt{2}g\vert a_0\vert^2\\
&0=\left(-i\Delta_c-\dfrac{\kappa}{2}\right)a_0+i\sqrt{2}ga_0q_0+\Omega_d\\
&0=\left(i\Delta_c-\dfrac{\kappa}{2}\right)a^{\dagger}_0-i\sqrt{2}ga^{\dagger}_0q_0+\Omega_d,
\label{eq:omitqlefirstorder}
\end{split}
\end{equation}
and 
\begin{equation}
\begin{split}
&-i\delta q_+=-\dfrac{\gamma_m}{2}q_+-\omega'_m p_+\\
&-i\delta p_+=-\dfrac{\gamma_m}{2}p_+-\omega'_m q_++\sqrt{2}g(a^{\dagger}_0a_++a_0a^{\dagger}_+)\\
&-i\delta a_+=\left(-i\Delta_c-\dfrac{\kappa}{2}\right)a_++i\sqrt{2}g(a_+q_0+a_0q_+)+1\\
&-i\delta a^{\dagger}_+=\left(i\Delta_c-\dfrac{\kappa}{2}\right)a^{\dagger}_+-i\sqrt{2}g(a^{\dagger}_+q_0+a^{\dagger}_0q_+).
\label{eq:omitqlesecondorder}
\end{split}
\end{equation}
For asymptotic steady state, we approximately consider $p_0=0$, i.e., the displacement dissipation $-\gamma_mq_{0,+}/2$ is neglected. Then based on above two sets of dynamic equations, the steady state solutions can be given finally by:
\begin{equation}
\begin{split}
&q_0=\dfrac{2g_0}{\omega_m}\vert a_0\vert ^2\\
&a_0=\dfrac{\Omega_d}{i(\Delta-\Delta')+\dfrac{\kappa}{2}}
\label{eq:omitsoulution}
\end{split}
\end{equation}
and 
\begin{equation}
\begin{split}
a_+=\dfrac{\left\{(\omega_m^2-\delta^2-i\delta\gamma_m/2)\left[-i(\Delta+\delta)+\dfrac{\kappa}{2}\right]+\beta\right\}}{\left(\omega_m^2-\delta^2-i\delta\gamma_m/2\right)\left[i(\Delta-\delta)+\dfrac{\kappa}{2}\right]\left[-i(\Delta+\delta)+\dfrac{\kappa}{2}\right]+i2\beta\Delta}.
\label{eq:solutionreslut2}
\end{split}
\end{equation}
Here Eq. \eqref{eq:solutionreslut2} are exactly the same with Eqs. \eqref{eq:solutionreslut1} in the main text.

\section{Derivation of the oscillator entanglement}
\label{Derivation of the oscillator entanglement}
Based on the Hamiltonian \eqref{eq:HamiltonianPTentna}, we have
\begin{equation}
\begin{split}
&\dfrac{d}{dt}\langle{b_1^{\dagger}b_1}\rangle=-iu\left(\langle{b_1^{\dagger}b_2}\rangle-\langle{b_1^{\dagger}b_2}\rangle^{*}\right)-\gamma\langle{b_1^{\dagger}b_1}\rangle+\gamma n_b\\
&\dfrac{d}{dt}\langle{b_2^{\dagger}b_2}\rangle=-iu\left(-\langle{b_1^{\dagger}b_2}\rangle+\langle{b_1^{\dagger}b_2}\rangle^{*}\right)+\gamma'\langle{b_2^{\dagger}b_2}\rangle-\gamma' n'_b\\
&\dfrac{d}{dt}\langle{b_1^{\dagger}b_2}\rangle=-\left(\dfrac{\gamma-\gamma'}{2}\right)\langle{b_1^{\dagger}b_2}\rangle-i\mu \left(\langle{b_1^{\dagger}b_1}\rangle-\langle{b_2^{\dagger}b_2}\rangle\right)\\
&\dfrac{d}{dt}\langle{b_1^2}\rangle=(-2i\omega-\gamma)-2i\mu\langle{b_1b_2}\rangle\\
&\dfrac{d}{dt}\langle{b_2^2}\rangle=(-2i\omega+\gamma')-2i\mu\langle{b_1b_2}\rangle\\
&\dfrac{d}{dt}\langle{b_1b_2}\rangle=\left(-2i\omega-\dfrac{\gamma-\gamma'}{2}\right)\langle{b_1b_2}\rangle-i\mu\left(\langle{b_1^2}\rangle+\langle{b_2^2}\rangle\right),
\label{eq:secondorderen}
\end{split}
\end{equation}
then the covariance matrix can be calculated by the transition matrix
\begin{equation}
S=
\begin{pmatrix}
 T & 0 \\  
 0 & T
\end{pmatrix}\,\,\,\,\,\,\,\,
where\,\,\,\,\,\,\,\,
T=\dfrac{1}{\sqrt{2}}
\begin{pmatrix}
 1 & 1 \\  
 i & -i
\end{pmatrix}
\end{equation}
Furthermore, the \textit{logarithmic negativity} can be calculated according to the following relationship  
\begin{equation}
E_n=\max[0, -\ln(2\zeta)].
\end{equation}
In this expression, $\zeta$ is the smallest symplectic eigenvalue of the partially transposed covariance matrix $\tilde{C}$ obtained from $C$ just by taking $p_j$ in $-p_j$. This symplectic eigenvalue can be achieved by calculating the square roots of the ordinary eigenvalues of $-(\sigma\tilde{C})^2$, where $\sigma=J\oplus J$ and $J$ is a $2\times2$ matrix with $J_{12}=-J_{21}=1$ and $J_{11}=J_{22}=0$.
\end{widetext}


\begin{thebibliography}{35}
\bibitem{shuoming} The accurate eigenvectors should correspond to $\lambda_\pm$ in Eq. \eqref{eq:eigenvalues}. Because $\gamma\ll\mu,\omega_m$, we approximately consider that $\lambda_\pm=\omega_m\pm\mu$ and neglect term $c^\dagger d+H.c.$ here.
\end{thebibliography}

\end{document}